\documentstyle[12pt,epsf]{article}
\def\Ex#1{\langle#1\rangle}
\def\qed{\nobreak\kern 1em \vrule height .5em width .5em depth 0em}
\def\vbar{\mathchoice{\vrule height6.3ptdepth-.5ptwidth.8pt\kern-.8pt}
   {\vrule height6.3ptdepth-.5ptwidth.8pt\kern-.8pt}
   {\vrule height4.1ptdepth-.35ptwidth.6pt\kern-.6pt}
   {\vrule height3.1ptdepth-.25ptwidth.5pt\kern-.5pt}}

\def\date
   {\noindent Date: October 10, 1996
    \medskip}

\newcount\subno      
\newcount\subsubno   
\newcount\appno      
\newcount\appeqno    
\newcount\tableno    
\newcount\figureno   
\def\nsection#1
   {
    \vskip0pt plus-.1\vsize\vskip24pt plus12pt minus6pt
    \subno=0 \subsubno=0
    \addtocounter{section}{1}
    \noindent {\bf \thesection. #1\par}
    \noindent}
\def\nsubsecnn#1
   {\vskip-\lastskip
    \vskip24pt plus12pt minus6pt
    \noindent {\bf #1\par}
    \medskip
    \noindent}
\begin{document}
\setcounter{page}{1}
\pagestyle{plain}
\setcounter{equation}{0}
%
%
%
\ \\[12mm]
\begin{center}
    {\bf LOCALISATION TRANSITION  OF A DYNAMIC REACTION FRONT}
\\[15mm]
\end{center}
\begin{center}
\normalsize
M.\ J.\ E.\ Richardson\\[2mm]
{ \it Department of Theoretical Physics\\ University of Oxford,\\
1 Keble Road, Oxford, OX1 3NP, U.K.}\\[4mm]
and\\[4mm]
M.\ R.\ Evans\\[2mm]
{\it Department of Physics and Astronomy\\
        University of Edinburgh\\ Mayfield Road, Edinburgh EH9 3JZ,
U.K.}
\end{center}
\ \\[8mm]
\noindent {\bf Abstract:} 
We study the reaction-diffusion process $A+B
\rightarrow \emptyset$ with injection of each species at
opposite boundaries of a one-dimensional lattice and bulk driving of each species
in opposing directions with a hardcore interaction. The system shows the novel feature of
phase transitions between localised and delocalised reaction zones as
the injection rate or reaction rate is varied. An approximate analytical
form for the phase diagram is derived
by relating both the domain of reactants $A$
and the domain of reactants $B$ to asymmetric exclusion processes
with open boundaries, a system for which the phase diagram is known exactly,
giving rise to three phases.
 The reaction zone width $w$ is described by a finite size scaling form relating the early time growth, relaxation time and
saturation width exponents. In each phase the exponents are distinct
from the previously studied case where the reactants diffuse isotropically. 
\\[12mm]
\noindent Published: \ \ {\it J. Phys. A } {\bf 30} (1997) 811--818
\\[3mm]
\rule{7cm}{0.2mm}
\begin{flushleft}
\parbox[t]{3.5cm}{\bf Key words:}Reaction-diffusion, reaction zone localisation, phase transitions, asymmetric  exclusion process.
\parbox[t]{12.5cm}{ }
\\[2mm]
\parbox[t]{3.5cm}{\bf PACS numbers:}    05.40+j, 64.60Cn, 82.20.Mj
\end{flushleft}
\normalsize
\thispagestyle{empty}
\mbox{}
\pagestyle{plain}
%
%
%
\newpage
\baselineskip=18pt plus 3pt minus 2pt
\setcounter{page}{1}
\setcounter{equation}{0}
\nsection{Introduction}
The behaviour of the two-species reaction-diffusion system $A+B
\rightarrow \emptyset$ has been widely
studied over the last decade\cite{general}. For a given realisation of this process,
choices must be made for the type of motion allowed for the $A$ and
$B$ reactants and the environment in which the reactions take
place. For the case of purely diffusive reactants, situations of
particular interest have been i) the evolution of a
system of initially segregated reactants on an infinite lattice
\cite{GR} and ii) the steady-state
behaviour on a finite lattice with fixed currents of the $A$ species
at the left boundary and the $B$ species at the right boundary
\cite{BNR}. In both cases a reaction zone
(a region where reactions occur with non-negligible
rate)
is formed between the two domains of reactants. The scaling of the
reaction zone has been studied within the mean-field approximation
\cite{GR,BNR} giving a good account of the behaviour in dimension two and
above \cite{KK}. However, in one dimension the confined geometry makes
diffusive mixing of reactants inefficient. This in turn leads to
significant density fluctuations which invalidate the mean-field
approach: distinct non-mean-field scaling behaviour is seen
\cite{CDC}--\cite{BHC}. It is also
known \cite{Jan,IKR} that in the case of homogeneous initial
conditions (both species are distributed randomly) the model with hard
core exclusion and biased diffusion of reactants is in a different universality
class to that of the reaction-diffusion system where the reactants
diffuse isotropically.

In the present work we show that
on the one-dimensional finite system the introduction
of hard-core exclusion between reactants together with
{\it opposed} bulk driving, i.e. $A$ particles move stochastically to the right
and $B$ particles to the left, produces the
novel phenomena of phase transitions when the reaction rate
and boundary injection rate are varied. In particular
we find transitions between a phase
where the reaction zone is localised (occupying a vanishingly small fraction
of the system in the thermodynamic limit) to phases where it
is delocalised (filling the whole system).

Let us first discuss the situation where no bulk drive is imposed: the reactants are not driven but just diffuse.
On the infinite system with initially
segregated reactants, the reaction zone width $w$ is expected to grow
(in the scaling regime) as $t^{\alpha}$. Mean-field theory
predicts $\alpha = 1/6$ \cite{GR} whereas one-dimensional simulations
and scaling arguments point to $\alpha = 1/4$
\cite{CDC,AHLS,LAHS,CD,Cornell,Krap}. In fact an RG calculation \cite{BHC}
suggests logarithmic corrections
$w \sim t^{1/4} ( \log t )^{1/2}$ which may explain the
slow convergence to $\alpha = 1/4$ found in several
simulations \cite{ALHS,CD,Cornell}.

In the complementary situation of a  finite lattice of size $L$
where reactant $A$ is injected at the left boundary (site 1)
and reactant $B$ is injected at the right (site $L$), the system
attains a steady state as $t \to \infty$.
The scaling of the reaction zone is predicted in mean-field theory
to depend on the injected flux $j$, reaction rate $\lambda$
and diffusion constant of reactants $D$ through
$w \sim  ( D^2/ (j\lambda) )^{1/3}$ \cite{BNR}. This is distinct from
the one-dimensional case where
scaling arguments \cite{CD,Krap} and an RG calculation \cite{LC}
suggest $w \sim (D/j)^{1/2}$.
Recently it has been shown
that when the wandering
as well as the intrinsic contribution to the reaction
zone width is taken into account,
RG and Monte Carlo simulations predict $w \sim (D/j)^{1/2} ( \log L)^{1/2}$ \cite{BHC}.

We now consider the effects of hardcore exclusion (though still with
no drift) with reactions occurring between particles on neighbouring sites.
On the infinite system the exclusion  is not expected
to affect the value of $\alpha$.
 However, one effect of  exclusion on the finite lattice
is that that current can no longer be fully controlled from outside
the system, since a reactant can only enter if a boundary
site is empty. If one attempts to inject reactants at a
finite rate the resultant current is in fact ${\cal O}(1/L)$.
To see this note that in the bulk the current is $D\frac{ d \rho }{dx}$
where $\rho$ is the concentration. Now, due to exclusion
the maximum value $\rho$ can take is one, therefore,
if we assume a reaction zone $w \ll L$ the concentration
gradient, and hence the current, is of order $1/L$.
From this  we deduce $w \sim 1/j^{1/2} \sim L^{1/2}$
confirming that the reaction zone is localised
i.e. $ w/L \to 0 $ as $  L \to \infty$.

In order to relate the behaviour on the the finite system
to the infinite system
with initially segregated reactants 
a finite size scaling form appears natural.
The early time growth of the reaction zone on the finite
system should be the same as that on the infinite system
since the effect of the boundaries will not have been felt:
the boundaries will only influence the dynamics after a time $t \sim L^z$. Therefore,
ignoring any logarithmic corrections,
we propose a finite size scaling form
\begin{equation}
w \sim t^\alpha f( t/L^z)
\label{fss}
\end{equation}
From the $t \to \infty$ limit discussed in the previous paragraph
we have the scaling relation $1/2 = \alpha z$.
When $\alpha$ takes its expected value of 1/4 this
implies $z=2$,  consistent with a diffusive motion of reactants.
We ran simulations (with hardcore exclusion)
and found agreement with the scaling form but with
$\alpha \simeq 0.29$. This is consistent with other simulations
where a slow approach to $\alpha =0.25$ has been found.

In the following we consider the effects of opposed bulk driving together
with exclusion and with boundary injection on the reaction zone behaviour. We shall show
$i$) that phase transitions occur as the reaction rate and injection
rate are varied, and $ii$) in all phases  finite size scaling
forms (\ref{fss}) hold but have different
$\alpha, z$ to the isotropic diffusion case. In particular two phases are predicted
where
the reaction zone width varies as $w\sim {\cal O}( L)$ and is thus delocalised.
\nsection{Definition of the Model}
The model consists of a one-dimensional lattice of $L$ sites, each of
which can be occupied by a single particle, either of
type $A$ or $B$, or unoccupied (which we denote by 0). Type $A$ particles are injected at the left edge
of the system and type $B$ particles are injected at the right, both
with rate $\delta$. Once in the lattice, the $A$s always hop
to the right and the $B$s always hop to the left. If ever an $A$ particle and
a $B$ particle are on adjacent sites they react
with rate $\lambda$ and are both removed from the lattice.
Since particles cannot occupy the same site, we see that
once an $A$ $B$ pair are adjacent they simply wait
a random time, exponentially distributed with mean $\lambda^{-1}$,
before annihilating.
 An inert
product of the reaction is then deposited on the
sites previously occupied by the particles. However, this product
plays no further role in the evolution of the system. 

The dynamics can be written in terms of microscopic rules for
a bond of two adjacent sites $i,i+1$ in the bulk of the system
\begin{eqnarray}
A_{i}\ 0_{i+1} \rightarrow 0_{i}\ A_{i+1} &\mbox{and}& 0_{i}\ B_{i+1}
\rightarrow B_{i}\ 0_{i+1}\;\; \mbox{ both with rate $1$} \nonumber\\
A_{i}\ B_{i+1} &\rightarrow& 0_{i}\ 0_{i+1}\;\; \mbox{ with rate $\lambda$} 
\end{eqnarray}
where these rules are valid for $1 \leq i \leq (L-1)$. At the
boundaries we have injection of particles
\begin{eqnarray}
0_{1} \rightarrow A_{1} &\mbox{and}& 0_{L} \rightarrow B_{L}\;\;
\mbox{ both with rate $\delta$, at the lhs and rhs respectively.} 
\end{eqnarray}
(If either an $A$ or a $B$ particle ever reaches the opposite end of
the lattice it is removed with unit rate. This
seldom occurs and has a negligible effect on the behaviour of
the system.) 

The rules thus specified allow one to simulate the model numerically
by Monte Carlo techniques. We  define $a_i$ and $b_i$ as the binary
variables denoting occupation of site $i$ by an $A$ or $B$ particle
respectively (i.e. if site $i$ has an $A$ particle on it $a_i=1$
otherwise $a_i=0$). We also define $\Ex{X(t)}$ as the quantity $X$
at time $t$  starting from some fixed initial condition
and averaged over  the stochastic
dynamics up to time t. In a Monte Carlo simulation this
is conveniently performed by averaging 
over many systems, each one having been evolved
from the same initial state at $t=0$. Quantities of interest can now
be considered such as the particle densities $\Ex{a_i},\Ex{b_i}$.

As the dynamics does not allow for particles to pass through each other,
the system is segregated into two domains: a domain comprising $A$s and $0$s and
a domain comprising $B$s and $0$s. The {\it reaction front} is defined
as being at the instantaneous position of the interface of the two domains (strictly only defined when an $A$ and $B$ particle are on
adjacent sites). Due to the fluctuating particle currents flowing into the interface, the reaction front will move
stochastically through the lattice. The probability of finding the reaction 
front at bond
$i,i+1$ at time $t$ is $\Ex{a_{i}(t)b_{i+1}(t)}$ giving the rate of production of the inert
product at site $i$, $R_i(t)$, as
$\lambda(\Ex{a_{i-1}(t)b_{i}(t)}+\Ex{a_{i}(t)b_{i+1}(t)})$. The
distribution $R(t)$ will be used to define the {\it reaction zone}, the
region where reactions can occur with non-zero rate, which in turn
characterises the phase structure of the model.

\nsection{The Phase Diagram}
The parameter space
of the model is spanned by   $\delta$ (the
injection rate) and $\lambda$ (the reaction rate). Using Monte Carlo
simulation with initial conditions of segregated
reactants, we examined the density profiles 
$\Ex{a_i},\Ex{b_i}$  over a time-scale less than the typical time
for large fluctuations of the reaction front to occur (short time average)
and 
in the $t\rightarrow\infty$ limit (long time average).
 The normalised distribution of
the inert product (proportional to the production rate) was also
measured. Considering
$0 \leq \delta \leq 1$ and
$0 \leq \lambda < \infty$ we determined three distinct phases, existing in the
following approximate regions (see figure 1):\\

 {\bf Phase I}.  $0<\delta<0.5$ and $\delta<<\lambda$. The
short-time average shows flat profiles of particle density $\simeq\delta$ away
from the reaction zone. The long-time average shows linearly decaying
profiles with a broad distribution of inert product. This implies the
reaction front is likely to visit all sites of the system:
 the reaction front is delocalised and the reaction zone
fills the system.
(The structures near the boundaries
are of finite lateral extent and are therefore unimportant in the large $L$
limit).

{\bf Phase II}.  $0<\lambda<1$ and $\lambda<<\delta$. Again, the
short-time average shows flat particle profiles away from the reaction zone, though in this phase
the density is a function of $\lambda$. As in phase I, the long-time
average shows clear evidence for a delocalised reaction front and a
reaction zone that fills the system. 

{\bf Phase III}. $\delta>0.5$ and $\lambda>1$. The short
and long time averages
show power law decays of the  density profiles
away from the boundaries and reaction front (i.e. long range structure). The
long-time average shows a peaked distribution of inert product near
the centre of the system. The reaction front is localised and the
width of the reaction zone is much less than the system size.\\

An approximate analytical form for the phase diagram can be easily
derived from a knowledge of the Totally Asymmetric Simple Exclusion
Process (TASEP). The
 TASEP   
comprises a system of particles moving  stochastically in 
a preferred direction  with hardcore exclusion
interactions on a lattice of $N$ sites.
In the case of open boundary conditions \cite{DEHP,SD,DEMall}
the microscopic rules are: $A_{i}\ 0_{i+1} \rightarrow 0_{i}\ A_{i+1}$
with rate $1$ across all sites $i,i+1$ where
$i=1\rightarrow{(N-1)}$ and at the boundaries there is injection of
particles $0_{1}
\rightarrow A_{1}$ with rate $\delta$ at the lhs and ejection of
particles $A_{N}
\rightarrow 0_{N}$  with rate $\beta$ at the rhs. The TASEP has 3
phases: a low density phase (density $\rho=\delta$) for $\delta<0.5$ and $\beta>\delta$, a high
density phase ($\rho=1-\beta$) for $\beta<0.5$ and $\delta>\beta$ and a maximal current
phase ($\rho = 0.5$) for $\delta>0.5$ and  $\beta>0.5$.

A simple approximation to the reaction diffusion model
can be made by considering each domain of reactants
as a TASEP (hopping to the right for the $A$ particles
and to the left for the $B$ particles).
The underlying assumption is a separation of time-scales
i.e. that the domains equilibrate to TASEP
profiles faster than
the time taken for large fluctuations in the position
of the reaction
front to occur. Therefore, an effective ejection rate for the $A$ particles can
be defined as $\beta_{eff}\simeq\lambda\rho_{B}$ where $\rho_{B}$ is the
density of $B$ particles at the reaction front. As we are
assuming the reaction front is stationary on the time-scale of the
TASEP $\rho_{B}\simeq\rho_{A}$. In the TASEP  $\beta$ can be
thought of as the density of a reservoir of holes just outside the rhs
of the
lattice, implying $\beta_{eff}\simeq(1-\rho_{A})$. Hence,
a self-consistent relation can be written for
 $\beta_{eff}$ in terms of $\lambda$:
\begin{eqnarray}
\lambda(1-\beta_{eff}) &=& \beta_{eff}.
\label{beff}
\end{eqnarray}
With this relation the phases of the TASEP can be
reinterpreted in terms of the reaction-diffusion system. Defining
$\rho$ as the (short-time average) density of reactants far away from the
reaction front, we find the following phase boundaries
and densities within this approximation: \\
\newline
{\bf Phase I} \ \ in the region $0<\delta<0.5$ and
$\delta < \frac{\lambda}{1+\lambda}\; , \; \; \rho=\delta\;$.\newline 
{\bf Phase II} \ in the region
$0 < \lambda < 1$ and $\frac{\lambda}{1+\lambda}< \delta \; ,\;\;
\rho=(1+\lambda)^{-1}\;$.\newline
{\bf Phase III} in the region $\delta > 0.5$ and
$\lambda > 1 \; , \;\; \rho=1/2$.\\
\newline
In phase I, the TASEP approximation for the density of reactants is in excellent
agreement with simulation. However, in phase II $\rho$ is slightly
less than $(1+\lambda)^{-1}$, the TASEP prediction. It also appears
from simulations,
 that the phase II/phase III phase boundary
may be displaced from its predicted position
(although 
the usual difficulties of locating a second order phase transition
make it hard  to be sure).
These observations suggest that the
above approximation for $\beta_{eff}$ might be
improved: the approximation clearly neglects fluctuations in the
position of the reaction front as well as correlations in time for
the occurrence of reactions.

\nsection{Localisation Measurements}
A quantity of physical relevance - the total amount of inert product
deposited onto a site up to time $t$,
${\cal{R}}_i(t)=\int_0^tR_i(t')dt'$, can be used to study the extent to
which the reaction front is localised in each of the
phases. Therefore, the time-dependent width
of the distribution of deposited inert product,
which we take to define the reaction zone, is given by:
\begin{eqnarray}
w^2(t) &=&
\sum_{k=1}^{L}(k-k_f)^2{\cal{R}}_k(t)/\sum_{k=1}^{L}{\cal{R}}_k(t) \mbox{
where } k_f=\sum_{k=1}^{L}k{\cal{R}}_k(t)/\sum_{k=1}^{L}{\cal{R}}_k(t)
\label{width}
\end{eqnarray}

\noindent{\bf Steady State Width.} For the reaction front to be
considered
 localised, the steady state
width, $w_{\infty}=\lim_{t\rightarrow\infty}w(t)$,
must scale as $L^{\gamma}$ with $\gamma<1$. By Monte Carlo
simulation, the exponent $\gamma$ was measured for various $L$
 for the three phases, figure 1. The exponents
measured were: phase I $\gamma\simeq1.01\pm0.02$, phase
II $\gamma\simeq1.04\pm0.05$ and phase III $\gamma\simeq0.75\pm0.01$,
suggestive of $\gamma = \ 1, 1, 3/4$ respectively. In phase I and II the distribution of the inert product remains
constant in the limit $L\rightarrow\infty$. However, in
phase III the fraction of the system containing the inert product vanishes as $\sim{L^{-1/4}}$ i.e. the
reaction front is localised in the middle of the system.\\

\noindent{\bf Relaxation Time.} In order to investigate the dynamical
properties of the model we considered the time taken for the system to relax
to steady state behaviour (expected to vary as $\tau\sim{L^z}$).  We measured
the exponent $z$ by evolving the
system from an initial configuration with the reaction front at a
boundary, i.e. the system full of $A$ particles, and examining the $L$
dependence of $\tau$ the time taken for the reaction front to reach the centre of the
system, figure 2. Monte Carlo simulations averaged over $\sim500$
systems gave: phase I $z\simeq1.93\pm0.08$, phase II
$z\simeq2.02\pm0.05$ and phase III $z\simeq 2.25\pm0.02$,
suggestive of $2, 2, 9/4$ respectively.
In phase I and II simulations suggest
that the reaction front behaves as an unbiased random
walker: after waiting a time $\sim{L^{2}}$ the reaction front will have explored the
whole system. In the TASEP approximation this is expected from the flat
profiles and the lack of long-range correlations in these phases.
Indeed, in the limit $\lambda \to 0$ (in phase II)
the system will be devoid of zeros except when occasional reactions
happen and it is easy to show that the dynamics of the front
is exactly that of a random walk. 
In phase III the time taken for the the front to reach
the centre of the system varies as $\sim{L^{9/4}}$, i.e. the
motion is subdiffusive. However, when the time taken for the front to
travel from the centre of the system to a boundary was measured it was
found to increase exponentially with $L$. This
implies that in phase III the reaction
front is stable over short periods of time, but the net motion is
biased towards the centre of the system.\\

\noindent{\bf Finite Size Scaling Forms.} A commonly studied situation in reaction-diffusion systems is the
evolution of the system from an initial configuration (at $t=0$) of
segregated reactants. In this case, initially half the lattice
($i=1\rightarrow{L/2}$) contains $A$ particles with density $1/2$ and
the other half of the lattice ($i=L/2\rightarrow{L}$) contains $B$
particles, also of density $1/2$. We expect the growth of $w(t)$ from
the initial condition $w(0)=0$ to follow a finite size scaling form
(\ref{fss}). As $w(\infty)=L^{\gamma}$ we have the scaling relation
$\gamma=z\alpha$ giving, in conjunction with our
above results, $\alpha=1/2$ for phases I and II and for
the localised phase III $\alpha=1/3$. 
The function $w(t)$ was measured for various $L$ for phases I,II and
III. The data collapse (figure 3) supports the following scaling forms
where $g_{I,II}, \tilde{g}_{I,II},h , \tilde{h}$ are scaling functions:
\begin{eqnarray}
\mbox{For phases I and II }\;\; w(t) =
t^{1/2}g_{I,II}\left(\frac{t}{L^{2}}\right)\;\;&
\mbox{or}&\;\; w(t) =
L \tilde{g}_{I,II}\left(\frac{t}{L^{2}}\right)\;. 
\label{fssIandII} \\
\mbox{For phase III }\;\; w(t) =
t^{1/3}h\left(\frac{t}{L^{9/4}}\right)
\;\;&\mbox{or}&\;\; w(t) =
L^{3/4}\tilde{h}\left(\frac{t}{L^{9/4}}\right)\;.
\label{fssIII}
\end{eqnarray}
For phase II and III we believe the data collapses
are convincing evidence of the scaling form. 
However, for phase I larger simulations are really required to
confirm
the  early time growth region
($w \sim t^{1/2}$).

\nsection{Discussion}
We have introduced a variant of the $A+B\rightarrow\emptyset$
reaction-diffusion system which shows transitions between phases with
delocalised and localised reaction zones as the injection rate
$\delta$ and reaction rate $\lambda$ are varied. By treating a single domain of a
reactant as a Totally Asymmetric Exclusion Process (TASEP), an
approximate relationship (\ref{beff}) for an effective `ejection
rate', $\beta_{eff}$, can be
found in terms of  $\lambda$ , allowing an
approximate phase diagram to be derived.
A particularly interesting transition is that from phase I (delocalised
reaction zone)
to phase III (localised reaction zone), which occurs (for $\delta <1/2$)
by simply increasing the reaction rate.

In each of the phases the exponents \{$\alpha,\; z,\; \gamma$\} (early time growth of the
reaction zone $w\sim t^{\alpha}$; relaxation time $t\sim L^{z}$;
reaction zone saturation width $w\sim L^{\gamma}$) obey the finite size
scaling relation $\alpha z=\gamma$. In each case, \{1/2,\ 2,\ 1\} for the delocalised
phases and \{1/3,\ 9/4,\ 3/4\} for the localised phase, these exponents
are distinct from those of the model
discussed in the introduction where the reactants have no bias (\{1/4,\ 2,\ 1/2\})
 suggesting a different universality class for the
present model. 
We believe that the new exponents are related to the
behaviour of current fluctuations in the TASEP 
\cite{DEMall} which in turn are related to KPZ exponents,
although we could not find a clean argument.

It would be interesting to examine the robustness of the model to {\it i})
relaxation of the bias to allow for a small backward hopping rate of
reactants, and {\it ii}) relaxation of the hardcore exclusion between the
different species. If the TASEP approximation is valid the effect of
relaxing the bias should preserve the phase structure of the model:
partially asymmetric exclusion and asymmetric exclusion belong to the same
universality class. However, the relaxation of 
hardcore exclusion between $A$ and $B$ reactants may strongly
affect phase II where the reaction rate is low. 

Lastly it would be illuminating to make precise the relation of our 
TASEP approximation of the reaction diffusion system
to a conventional mean field theory which would
be expected to hold in dimension $d \geq 2$. One could then
explore the feasibility of the RG approach which has been successful
in the case where the reactants diffuse isotropically \cite{BHC}

\nsubsecnn{Acknowledgments}
We thank J. Cardy,  S. Cornell, M. Howard, Z. R\'acz and S. Redner for useful
discussions and correspondence.
MJER acknowledges financial support from EPSRC
under Award No. 94304282.
MRE is a Royal Society University Research Fellow.

%
%


\newpage
\nsubsecnn{Figure Captions}
{\bf Figure 1.} \newline
Short and long time averages for each phase
from Monte-Carlo simulation. The 
filled squares plot the density of $A$; the empty squares plot the
density of $B$; the line plots the distribution of inert
product, normalised so that maximum value is 1/2. Long time averages
were carried out over $10^8$ MCS. The short time averages were
over 300 MCS for phase I and 500 MCS for phases II and III
all averaged over 5000 systems. The values of
($\delta,\lambda$) are  Phase I (0.2, 9); 
Phase II (0.7, 0.3); Phase III (0.9, 10) .\newline
\\
{\bf Figure 2.} \newline
(a) Log-log plot of  the relaxation time $\tau$ (see main text)
versus system size $L$. The measured gradients
predict $\tau \sim L^z$ 
where  $z=1.93\pm0.08$ in phase I ; $z=2.02\pm0.05$ in
phase II ;  $z=2.25\pm0.02$ in phase III .\\
(b)
Log-log plot of the reaction zone width $w$ defined in
(\ref{width})  versus $L$. The measured gradients predict
$w \sim L^{\gamma}$ where
 $\gamma=1.01\pm0.02$ in phase I ; $\gamma=1.04\pm0.05$ in
phase II ; $\gamma= 0.75 \pm0.01$ in phase III \\
In both (a) and  (b) the values
 of ($\delta,\lambda$) are  phase I $(0.2, \infty)$; 
phase II $(1, 0.2)$; phase III $(1, \infty)$ .\newline
\\
{\bf Figure 3.} \newline
Finite size scaling collapses for various values of $L$ between 200
and 1000.
$w/L^{\gamma}$ versus $t/L^z$  with \\
(a) $\gamma = 3/4 ,\; z =9/4$ (see Eq. \ref{fssIII}) for phase III\\
(b) $\gamma = 1 ,\; z =2$ (see Eq.  \ref{fssIandII}) for phase I
(empty symbols) and phase II (filled symbols).
\newpage
\begin{figure}
\begin{minipage}[b]{5cm}
\epsfxsize 5 cm
\epsfysize 4 cm
\epsfbox{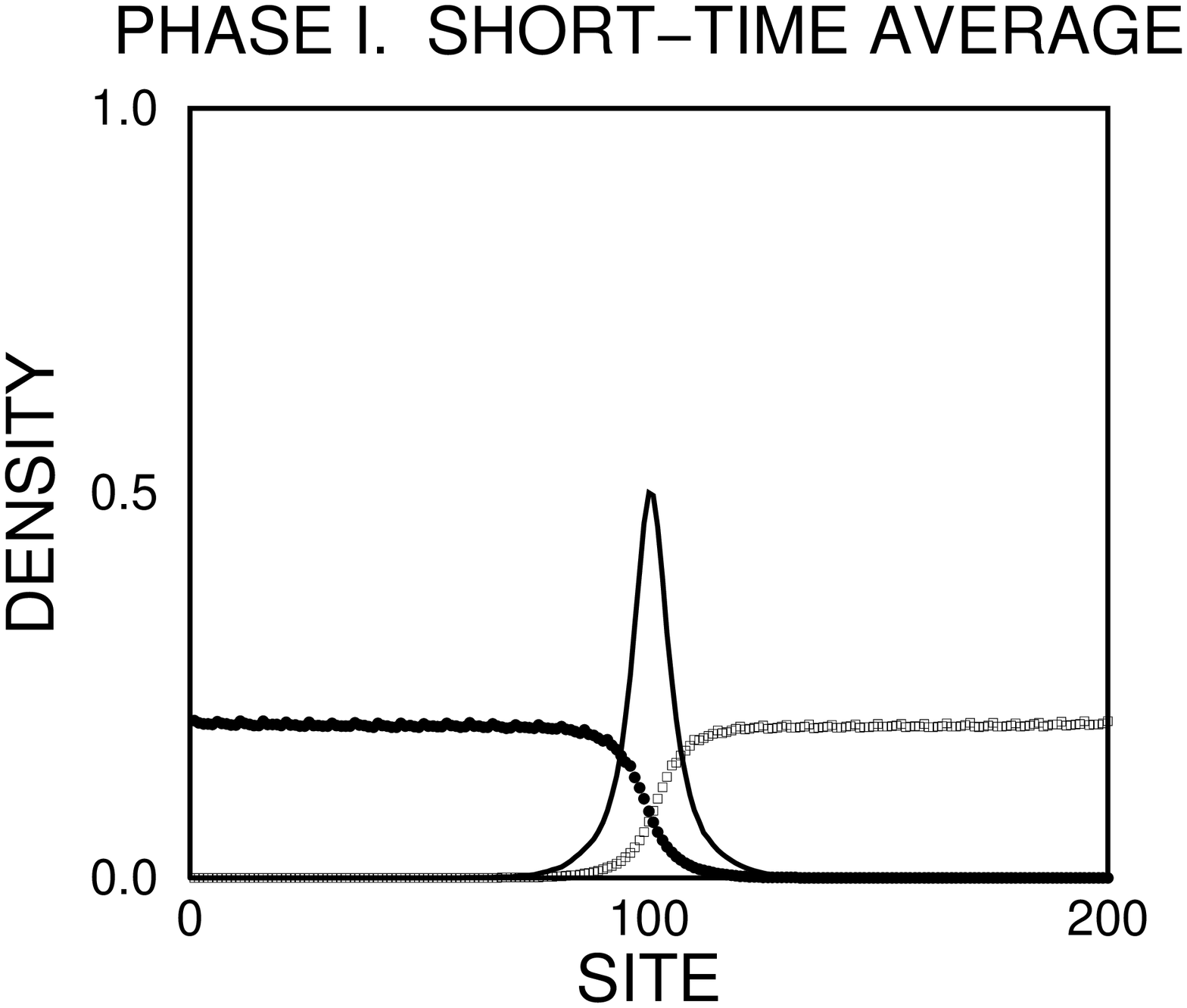}
\epsfxsize 5 cm
\epsfysize 4 cm
\epsfbox{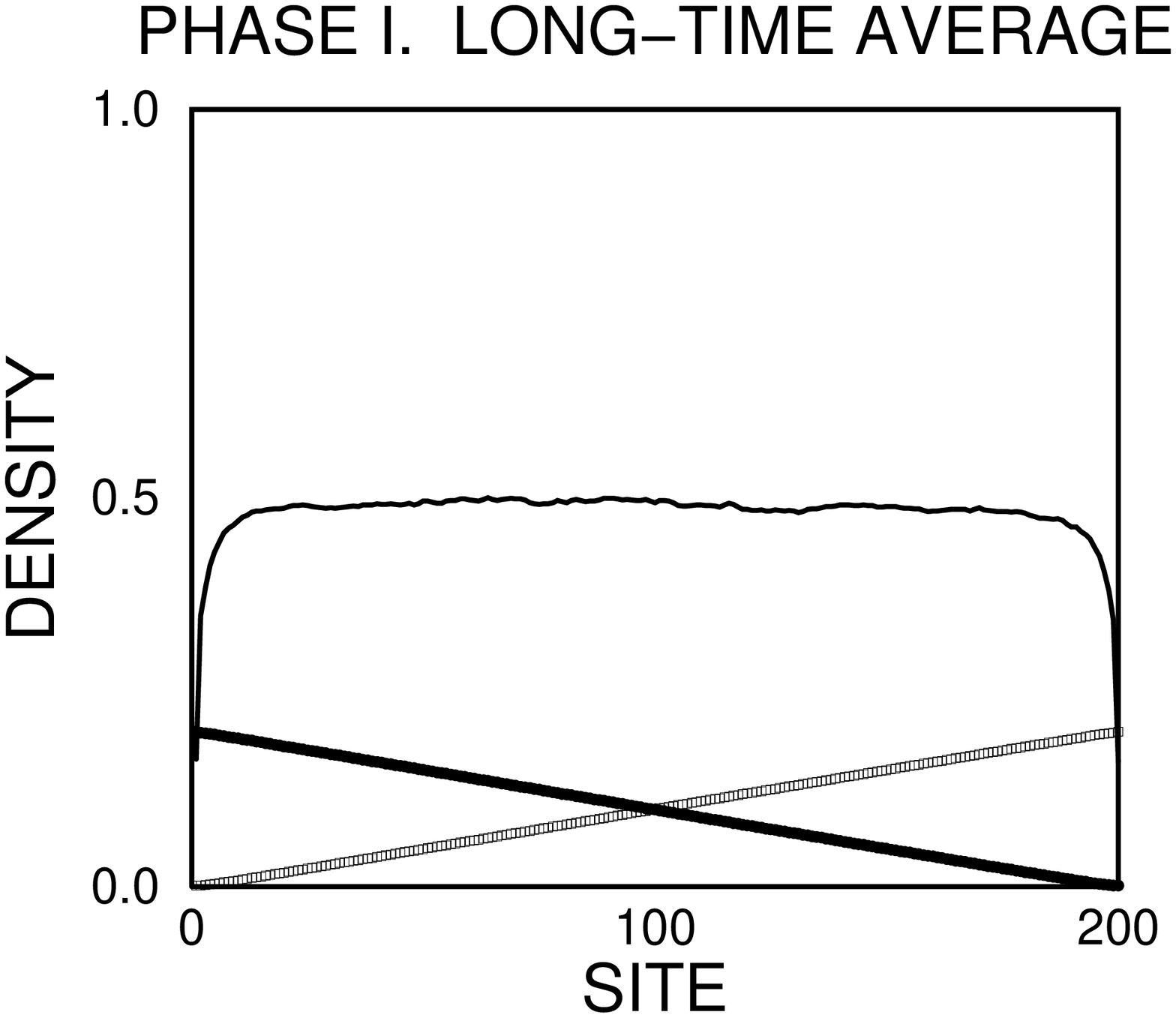}
\end{minipage}
\begin{minipage}[b]{5cm}
\epsfxsize 5 cm
\epsfysize 4 cm
\epsfbox{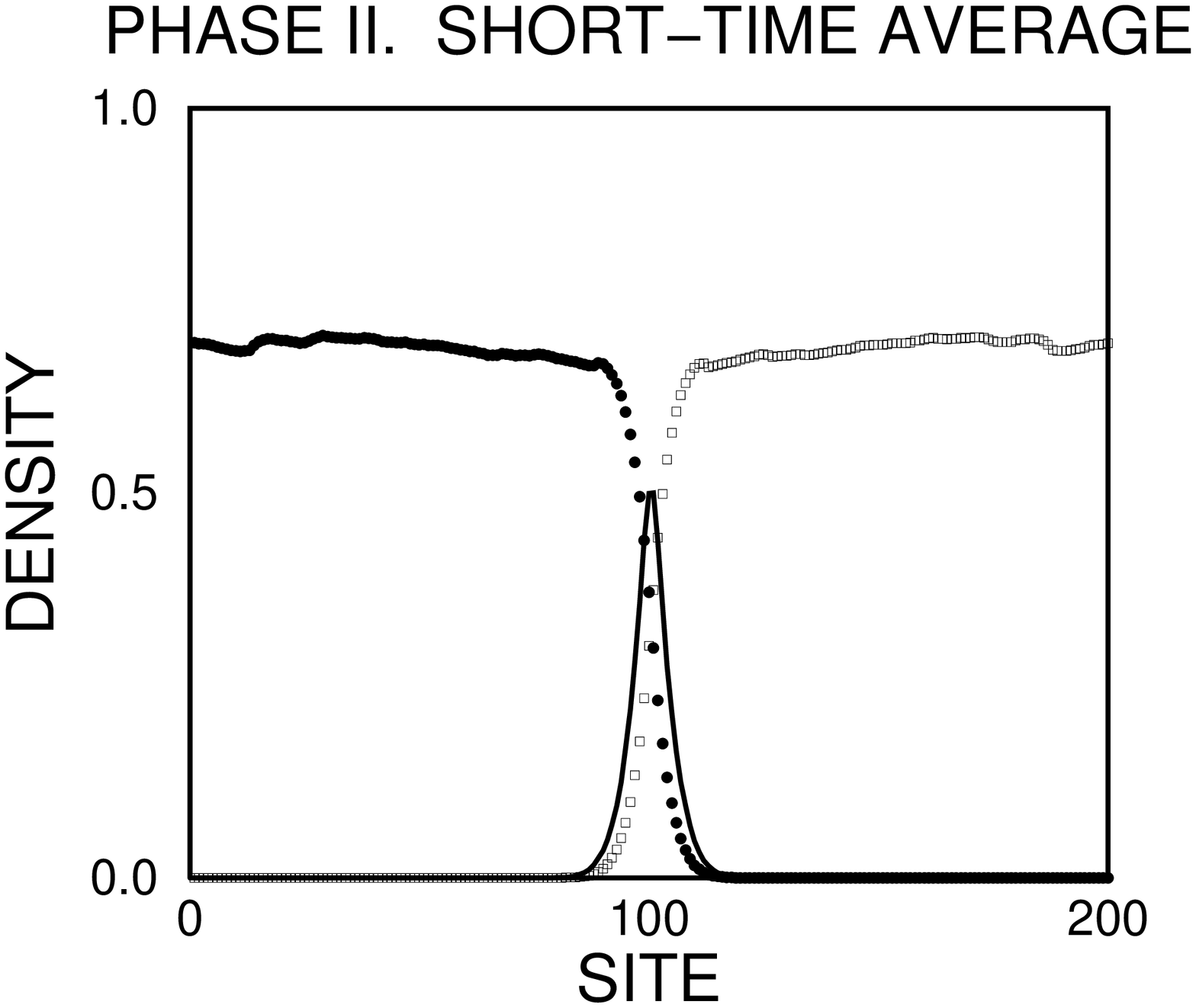}
\epsfxsize 5 cm
\epsfysize 4 cm
\epsfbox{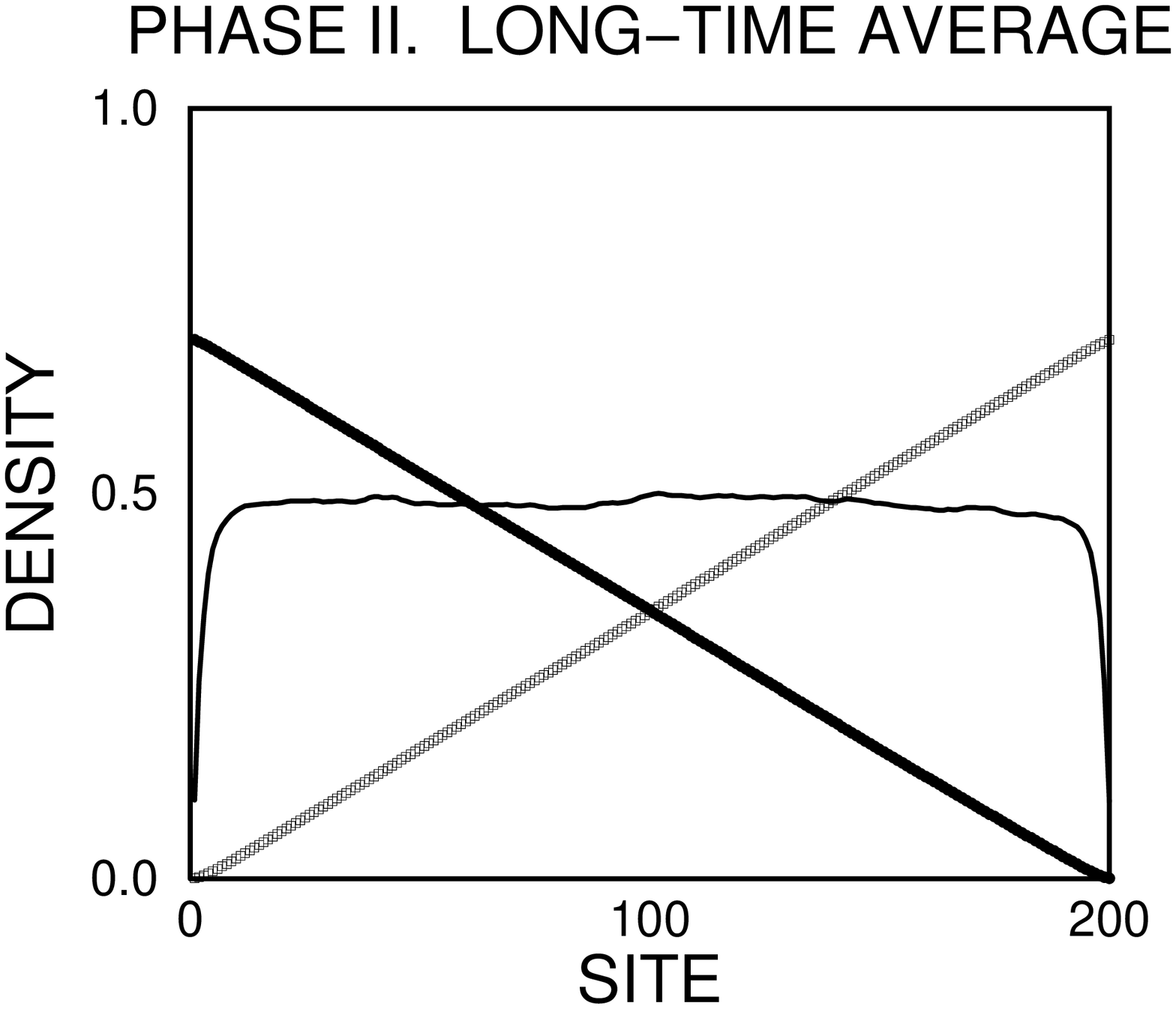}
\end{minipage}
\begin{minipage}[b]{4cm}
\epsfxsize 5 cm
\epsfysize 4 cm
\epsfbox{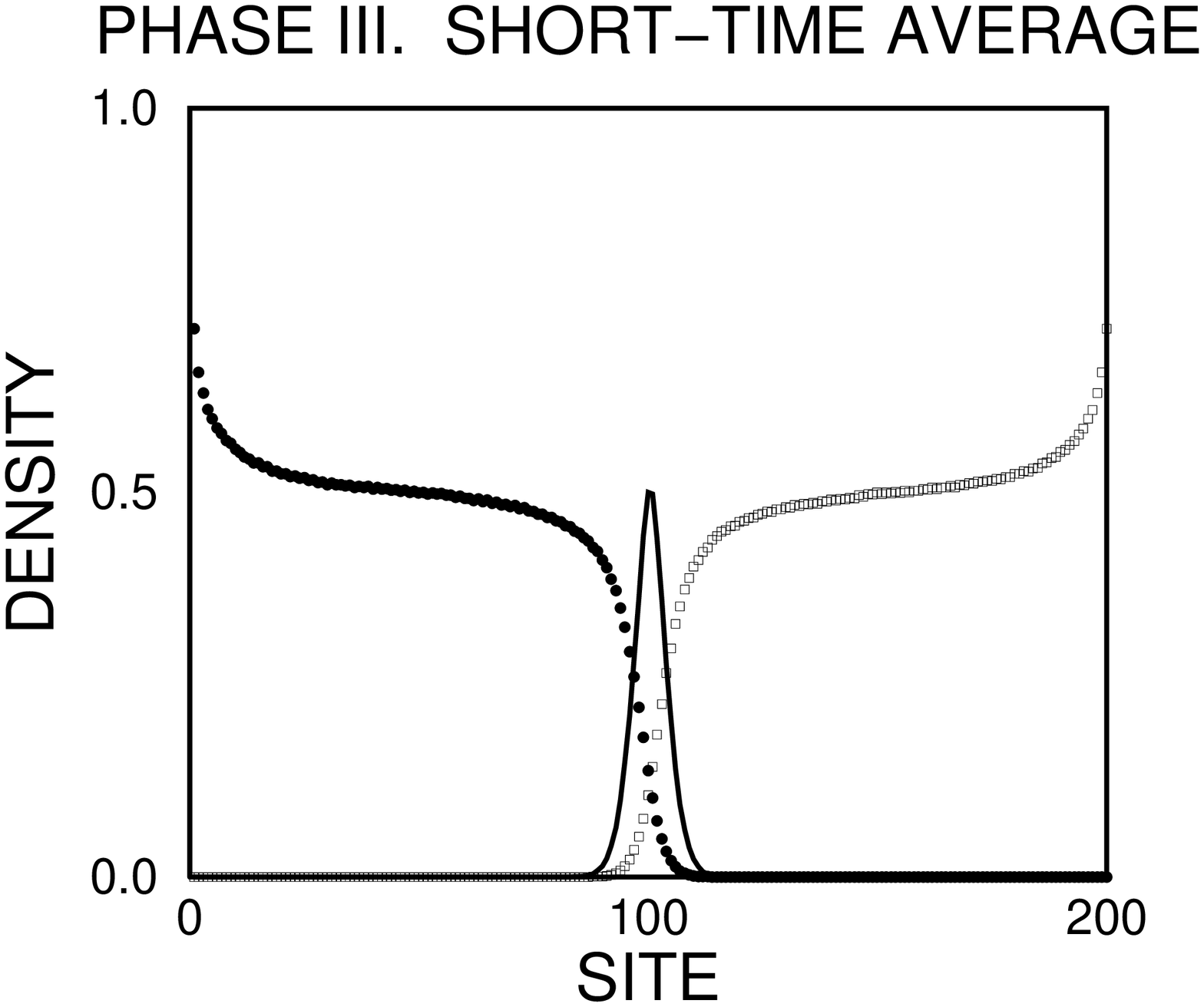}
\epsfxsize 5 cm
\epsfysize 4 cm
\epsfbox{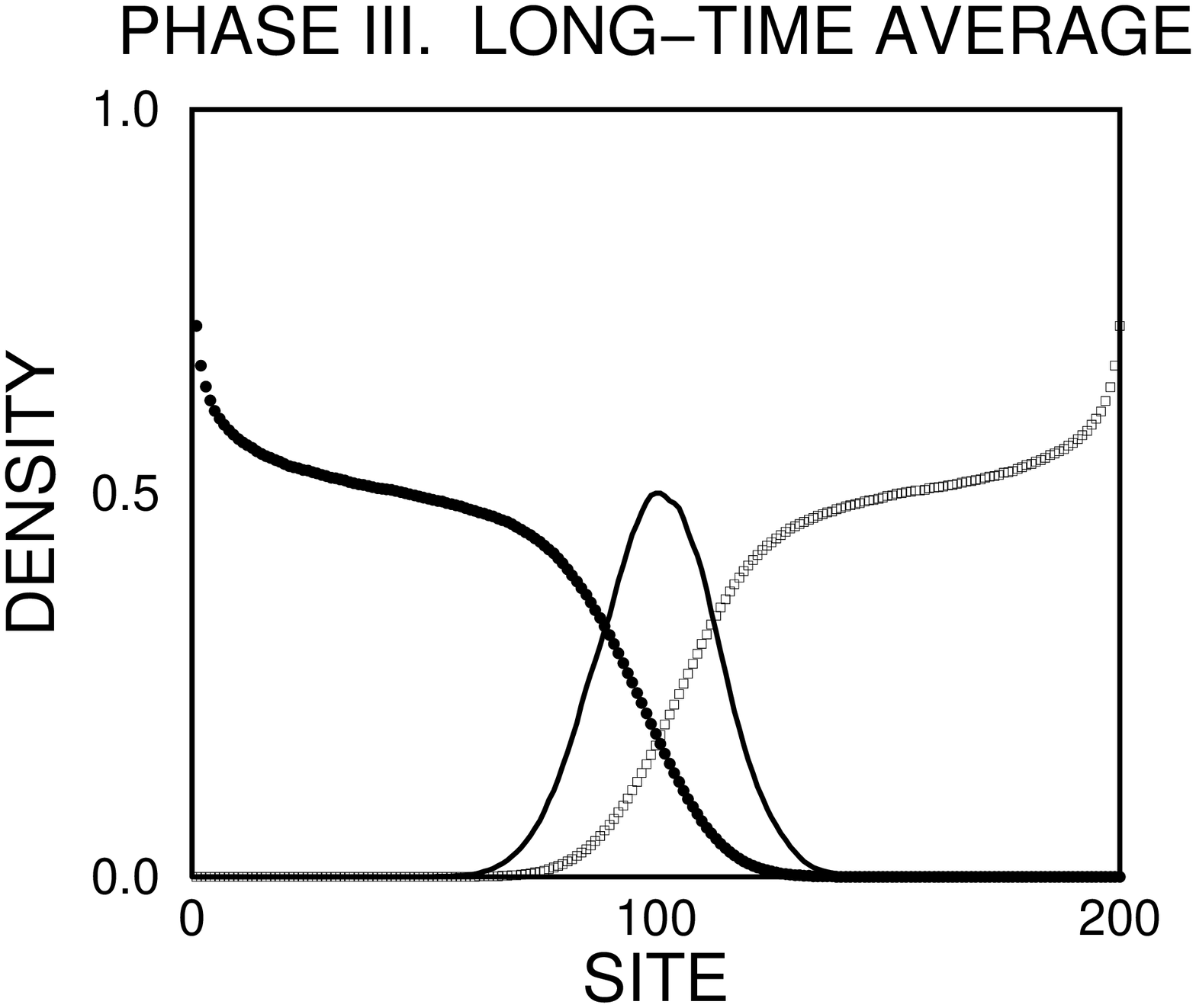}
\end{minipage}
\caption{}
\end{figure}

\clearpage
\newpage

\begin{figure}
\begin{minipage}[t]{8cm}
\epsfxsize 8 cm
\epsfysize 6 cm
\epsfbox{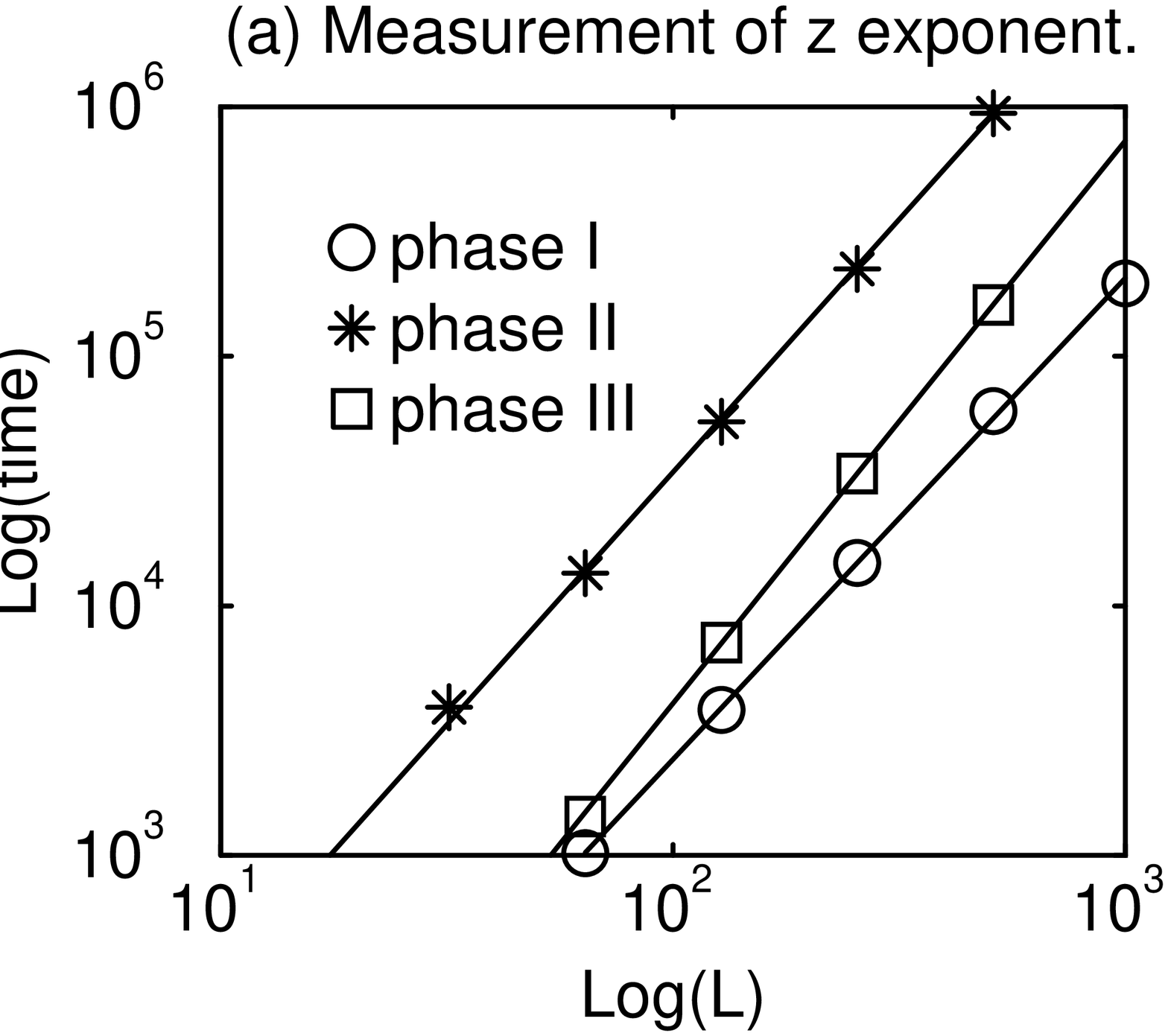}
\end{minipage}
\begin{minipage}[t]{8cm}
\epsfxsize 8 cm
\epsfysize 6 cm
\epsfbox{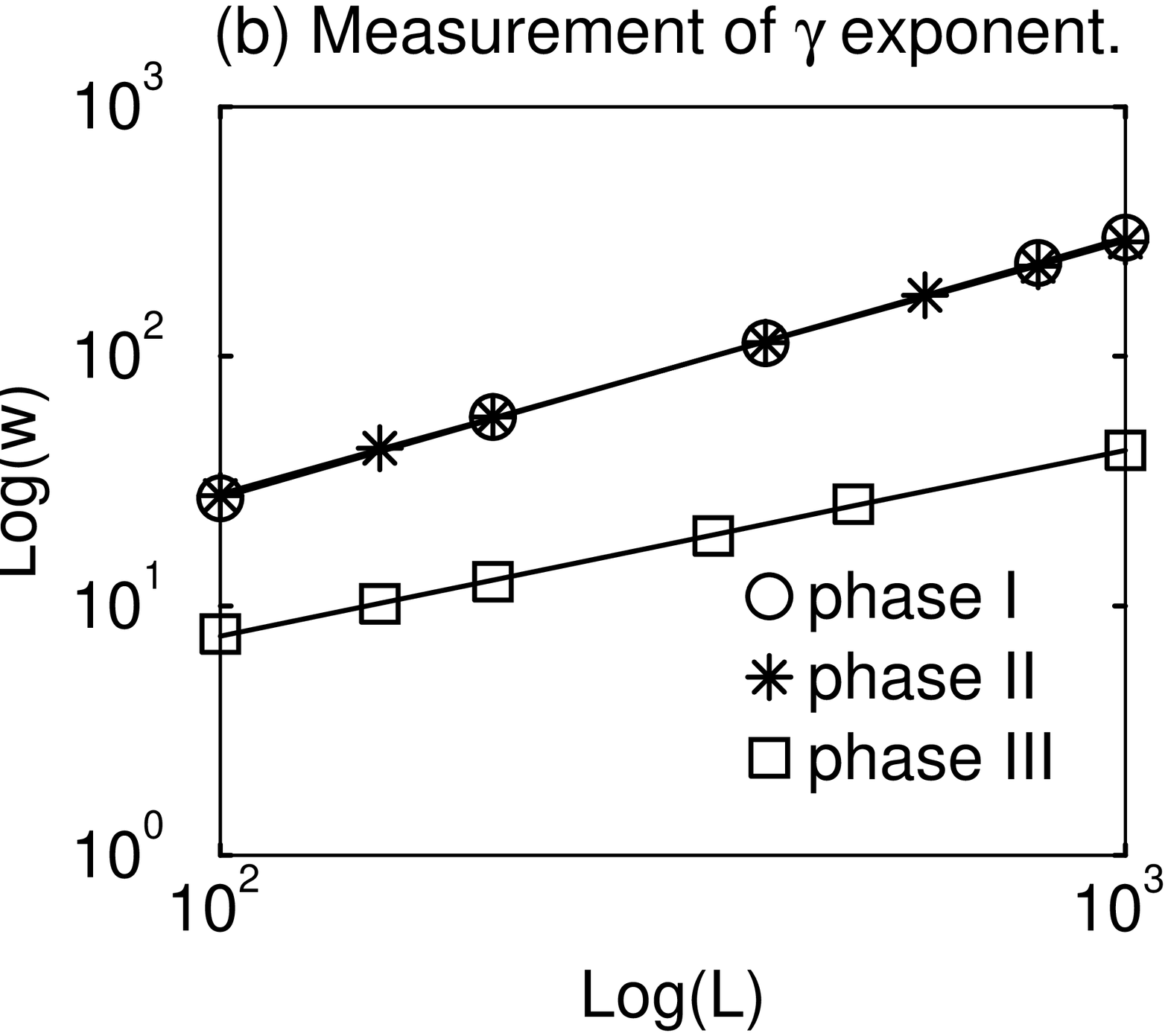}
\end{minipage}
\caption{}
\end{figure}

\clearpage
\newpage

\begin{figure}
\begin{minipage}[t]{8cm}
\epsfxsize 8 cm
\epsfysize 6 cm
\epsfbox{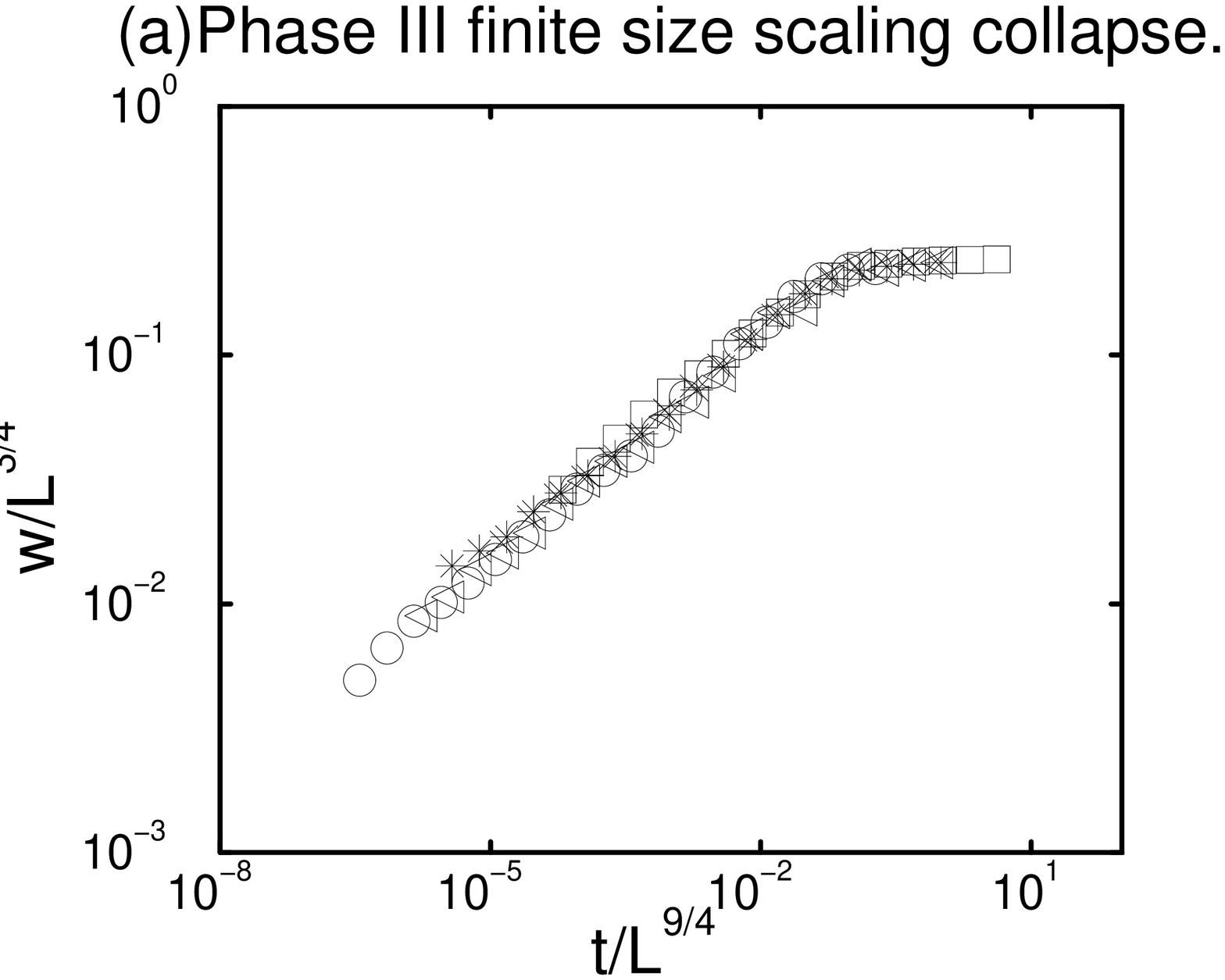}
\end{minipage}
\begin{minipage}[t]{8cm}
\epsfxsize 8 cm
\epsfysize 6 cm
\epsfbox{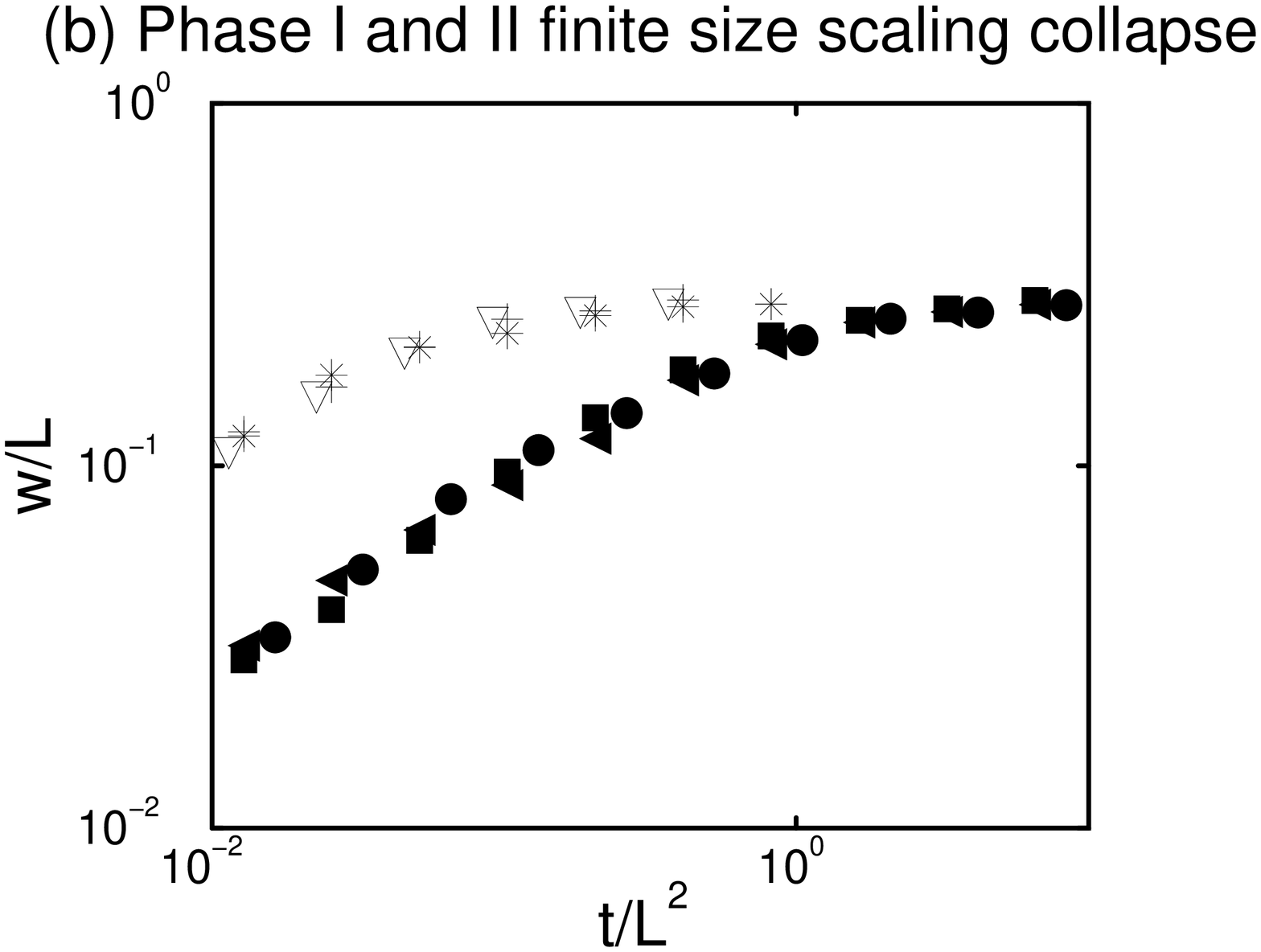}
\end{minipage}
\caption{}
\end{figure}

\end{document}